\documentclass[10pt,journal,compsoc]{IEEEtran}

\usepackage{dblfloatfix} 

\input{dependencies}

    \definecolor{ao(english)}{rgb}{0.0, 0.5, 0.0}
    
	\definecolor{armygreen}{rgb}{0.29, 0.33, 0.13}
\begin{document}
\title{Better Metrics for  
Ranking SE Researchers}

\author{George~Mathew,
        Tim~Menzies,~\IEEEmembership{Senior Member~IEEE}
\IEEEcompsocitemizethanks{\IEEEcompsocthanksitem George Mathew is a Ph.D candidate in   Computer Science at North Carolina State University. 
E-mail: george.meg91@gmail.com
\IEEEcompsocthanksitem Tim Menzies is a full professor in   Computer Science at North Carolina State University.
E-mail: timm@ieee.org
}
}

\markboth{IEEE Trans SE, ~Vol.~XXX, No.~XX, August~XXXX}%
{Mathew \MakeLowercase{\textit{\etal}}: Better Metrics for  
Ranking SE Researchers}

\IEEEtitleabstractindextext{%
\begin{abstract}
This paper
studies how SE researchers are ranked using   a variety of  metrics and data from  35,406 authors of
   35,391  papers from
34 top SE venues in the period  1992-2016.
Based on that analysis, we:
deprecate the widely used ``h-index'', favoring instead an alternate Weighted PageRank($\mathit{PR}_W$) metric  that is somewhat analogous to the
PageRank($\mathit{PR}$) metric  developed at Google.  Unlike the h-index, $\mathit{PR}_W$ rewards not just citation counts but also how often authors collaborate.
Using  $\mathit{PR}_W$,  we  offer  a  ranking  of  the  top-20 SE
authors in the last decade.

\end{abstract}
}

\maketitle

\IEEEdisplaynontitleabstractindextext

\IEEEpeerreviewmaketitle

\ifCLASSOPTIONcompsoc
\IEEEraisesectionheading{\section{Introduction}\label{sec:introduction}}
\else
\section{Introduction}
\label{sec:introduction}
\fi



 One metric that is commonly used
 to detect a ``good''  researcher is the number of articles they publish at leading
   venues (e.g. the h-index).   Such  citation-based metrics has been criticized as an incomplete 
   summary of the value of research. Many alternative schemes have been proposed but,
   so far, there is little agreement on which to use. 
   
   This paper studies the rankings of the  35,406 authors of
   35,391  papers from
34 top SE venues in the period  1992-2016. 
Various ranking schemes are proposed including
one method called 
$\mathit{PR}_W$ that rewards both citation counts and how often resarchers collaborate with each other.
As shown below, $\mathit{PR}_W$ is more trustworthy since it 
is {\em numerically stable} (defined below).
   Using that stable ranking, we offer a ranking of the top-20 SE authors in the last decade.

Overall, this paper makes the following contributions:
\be
\item We define  tests for a ``good'' metric of scholastic excellence: (a)~it should use information
from multiple sources (not   merely citation counts); (b)~the rankings offered by that metric are stable across minor
changes to its derivation parameters.
\item
We define an automatic test  for  metric numeric stability.
\item
We show that   $\mathit{PR}_W$ satisfies our tests for a ``good'' metrics.
\item
Using $\mathit{PR}_W$, we list the most high-profile SE researchers.
\item
We offer at \href{https://goo.gl/xnB6f3}{goo.gl/xnB6f3} all the data and scripts required to automatically repeat this analysis.
\ee
This last point is very important. While  prior studies have proposed methods to rank scholars in
software engineering, those methods had repeatability issues due to the subjective nature of some of the
decisions within that analysis.  Here, we seek   a ranking methods that is most repeatable since it is
most stable across a wide range of
subjective decisions.

The rest of this paper is structured as follows. After some preliminaries in the next section, we present the data used in this study.
This is followed by the results from that data that make us advocate for $\mathit{PR}_W$. Finally, using $\mathit{PR}_W$, we list
the top-ranked  authors in SE in the last decade.


Note that this paper is an extension to a prior study~\cite{mathew2017TSE}
that looked for topic trends in software engineering. That prior
work did not explore issues of author rankings, nor did it test
if different   ranking metrics resulted in different  author rankings.



\section{Preliminaries}
\subsection{Motivation}  
Why  is it important to study ranking metrics for software scholars?  We argue that just as software should be verified,
so too should this community verify the software models that  rank  SE scholars. For example, ranking   model includes derivation parameters which, if changed by small amounts, could potentially change the rankings of SE scholars. This paper verifies that our preferred ranking metric\footnote{According to Fenton~\cite{fenton14},   a ``measure'' is some
numeric value (e.g. ``h-index=28'') while
a ``metric'' is some combination of measure and threshold
(e.g. ``h-index over 20 is good'').  Nevertheless, we use the term ``metrics''
for all the different indicators studied here (h-index, $\mathit{PR}_W$ and two others)  since once they are   used
to rank ``N'' scholars, then that ``measure'' becomes a ``metric'' since it ranks scholars into worst, better, best.},  
$\mathit{PR}_W$, is stable across  a range of derivation parameters.

But taking a step backwards, why is it important to debate how this community recognizes scholastic achievement?
 Researchers are often  judged by their scientific contributions which helps them in their research and academic career.  A recent article in the Science Magazine~\cite{kuo17faculty} surveyed factors affecting tenure faculty hiring. They noted:
\begin{quote}
    \textit{In the tenure-track faculty job hunt, status counts.}
\end{quote}
 The survey suggests that the hiring committee at a research-intensive university valued most in an assistant professor candidate were the number of articles published in high-profile venues and the number of citations these articles receive. Teaching and service were deemed ``unnecessary credentials" and more often than not did not influence tenure selection~\cite{kuo17faculty}.
 The results, though not surprising, offer a reminder that with so many people vying for a limited
 number of  tenure-track faculty positions, ``trainees need to do more self-analysis of where they are and what the realities are for them to potentially become a faculty member''~\cite{kuo17faculty}.
 
Since status matters so much, it is wise to reflect on how that status is calculated and used to sort and select supposedly superior scholars.
We prefer the $\mathit{PR}_W$  metric,  for several reasons.

Firstly,
it use more information about an author; i.e. 
it rewards not just citation counts but also how often authors collaborate.

Secondly, 
how we measure our own community tells the world what we value most within this community.
Measures based on just  solo citations can encourage the belief that all that matters in research is individual success. However,
if our community decides to endorse collaboration-aware metrics, that say that SE researcher preferentially encourages a 
community
where researchers explore and assess and critique and improve each others' ideas.

Thirdly, as shown in this paper, this $\mathit{PR}_W$ metric is numerically stable.  
We say that a metric is {\em unstable} if small changes to its derivation parameters  lead to large deviations in the metric. Such instability is highly undesirable while ranking scholars, since small changes to how it is applied can lead to inappropriate
changes in the final ranking. In the Monte Carlo analysis reported below, we show that $\mathit{PR}_W$'s rankings are barely altered by perturbations 
to its derivation parameters.

Fourthly, as shown below, the rankings of scholars generated by these different metrics are not always the same.
However, $\mathit{PR}_W$'s rankings  most overlap with those from other metrics. Hence, adopting $\mathit{PR}_W$ will lead to least
future disputes about methods for ranking SE scholars.

\subsection{Ranking Metrics}
\label{sect:bg}

Over the years, success in SE has been defined and redefined by quite a few researchers in SE~\cite{ren2007automatic, fernandes2014authorship}. A few popular metrics to evaluate the success of an author are defined below

\textbf{Reputation Ranks}: Proposed by Ren \& Taylor~\cite{ren2007automatic}.  Authors are assigned weights based on the reputation of their affiliated institutions and the published venues. Authors are then ranked on this score.  While an insightful study, their calculations  were based on some  subjective decisions
by Ren \& Taylor. This makes it difficult to repeat, dispute, or improve on that study. Here, we seek to do better than Ren \& Taylor
by finding a ranking methods that is stable across a wide range of subjective decisions.

Fernandes in a 2014 article suggested four other metrics to evaluate author success
\bi
\item \textbf{Infl}: Total citations of the authors.  
\item \textbf{CoA}: Total number of co-authored articles.
\item \textbf{Frac}: Represents the fractional credit per author which is the cumulative citation count for each article weighted by the unit fractional credit. Unit fractional credit is the contribution of an author towards an article defined as the reciprocal of number coauthors in an article. 
\item \textbf{Harm}: Represents the harmonic credit per author which is the cumulative citation count for each article weighted by the unit harmonic credit. Unit harmonic credit(UHC) is the contribution of an author towards an article is defined as follows:
\ei

\[ \mathit{UHC \; for\;} i^{\mathit{th}} {\; \mathit{author}}\ = \frac{1/i}{1 + \frac{1}{2} + .. + \frac{1}{N}}\]
Note that some aspects of these metrics are problematic.  \textbf{Infl} and \textbf{CoA} do not account for co-authors and citations respectively. 
As to    \textbf{Harm}, this assumes that an author's position
in the author list precisely defines their contribution to a paper.
In the SE field, this may   not be case as documented in a recent
debate on this point\footnote{\url{https://goo.gl/A7kD8y}}.

\textbf{$h$-index}: Hirsch in 2005 proposed $h$-index~\cite{hirsch2005index} which is defined as the number of papers with citations greater than or equal to $h$. Although, this metric is a very popular metric used to represent an author's reputation, it fails to address some specific scenarios. For example, since $h$ index of an author considers only the number of citations of her article, it does not account for the prior work based on which the article is developed. For example,
consider the Google Scholar profile of Yann Gael Guenheneuc\footnote{\href{https://scholar.google.com/citations?user=_VV4cZYAAAAJ}{scholar.google.com/citations?user=\_VV4cZYAAAAJ}}. This author made many
highly-regarded contributions  (papers over 100 citations) for different domains in SE. This work
has influenced many other researchers to write their own,
highly cited articles. But, based on $h$-index, Guenheneuc ranks 50$^{th}$ amongst SE researchers. Thus, h-index fails to address scenarios that lead to greater contribution by other researchers.

\textbf{PR}: In this metric, authors are ranked using the decreasing order of their weighted PageRank. PageRank was initially developed by Page \etal~\cite{page1999pagerank} in 1999. 
\begin{equation}
\mathit{PR}_W(a_i) = 
(1- \theta) \frac{1}{N} + \theta\sum_{k \in N(a_i)} \frac{\mathit{PR}(a_k)}{|N(a_k)|}
\label{eq:pr}
\end{equation}
where $\theta$ is the probability of collaboration set between 0 and 1 and $N(a_i)$ represents the collaborators of an author $a_i$.
The authors of PageRank  argue that  PageRank is  a Markov Chain algorithm trying to stabilize the probability of transition between all the nodes (in this case authors) of a weighted graph of authors. 
A high value of $\theta$ leads to more faster convergence but can result in instability while a low value leads to slower convergence.
For much this paper we use $\theta = 0.5$, to allocate equal weight to both parts of PageRank (but see \tion{results3}
for studies where other values of $\theta$ are explored).

$\mathit{PR}_W$: Ding \etal propose a framework to modify PageRank to add a weight component~\cite{ding2009pagerank} to account for the individual contribution of an author.
The weighted PageRank of an author $a_i$ is defined as follows.

\begin{equation}
\mathit{PR}_W(a_i) = (1- \theta) \frac{W(a_i)}{\sum_{j=1}^N W(a_j)} + \theta\sum_{k \in N(a_i)} \frac{\mathit{PR}(a_k)}{|N(a_k)|} 
\label{eq:pr_w}
\end{equation}
Here, $W(a_i)$ represents the weight associated with an author. This weight can be number of citations, number of collaborators, h-index or any weighing factor. In our work we use either the number of publications (denoted {\bf $\mathit{PR}_{\mathit{publ}}$}) and number of citations
(denoted {\bf $\mathit{PR}_{\mathit{cite}}$}).
 
Note that
\textbf{PR} and \textbf{PR}$_W$ stress the value of collaborations 


\section{Data}
\label{sect:data}

The last section listed multiple metrics for ranking 
SE scholars:
{\bf PR,
Infl,
CoA,
Frac,
Harm,
h-in,
$\mathit{PR}_{\mathit{publ}}$},
or 
{\bf $\mathit{PR}_{\mathit{cite}}$}. Which should be used?

To answer that question,
we computed and compared rankings for 35,406 authors over a period of 25 years between 1992-2016. 
from 35,391 papers seen   from 34 SE conferences\footnote{RE,  ISSTA,  SSBSE,  ICST,  GPCE,  FASE,  ISSE,  FSE,  ASE,  ICSE,  SANER,  SCAM,  ICSM,  CSMR,
MSR, WCRE, ICPC, ESEM; 2012 to 2015.} and journals\footnote{REJ, TOSEM, TSE, ASEJ, IJSEKE, NOTES, JSS, SPE, IST, IEEEsoft. ESE SMR, SQJ, and STVR;
2012 to 2015.}.

This time period (25 years) was chosen since it encompasses recent trends in software engineering
such as the switch from waterfall to agile; platform migration from desktops to mobile; and the rise of cloud computing. Another reason to select this 25 year cut off was that we 
encountered increasingly more difficulty in accessing data prior to 1992; i.e. before the widespread use of the world-wide-web.

As to the list of venues, that was {\em initialized} using the top h5-index scores from Google Scholar, then {\em expanded}  after two rounds of input from
the SE community. More details on that process is extensively described by Mathew \etal~\cite{mathew2017TSE}.
What can be said here is that this venue list was explored for conclusion stability (specifically, the venue list was expanded until the $n+1$ expansion yielded the same conclusions as $n$). 
It should be noted that all the data collection method is automated. Thus, this list of venues could  expanded to additional venues in the future.

For studying and analyzing those venues we construct a database of 18 conferences, 16 journals, the papers published with the metadata, authors co-authoring the papers and the citation counts from 1992-2016. 
That data was collected in several  stages.

Firstly, for each venue,  DOI (Document Object Identifier), authors, title, venue \& year for every publication between 1992-2016 is obtained by scrapping the html page of DBLP\footnote{\url{http://dblp.uni-trier.de/}}. DBLP (\textbf{D}ata\textbf{B}ase systems \& \textbf{L}ogic \textbf{P}rogramming) computer science bibliography is an on-line reference for bibliographic information on major computer science publications. As of Jan 2017, dblp indexes over 3.4 million publications, published by more than 1.8 million authors. 

Second, for each publication, we obtain the corresponding citation counts using crossref's\footnote{\url{https://www.crossref.org/}} REST API.


Finally, since the data is acquired from three different sources, a great challenge lies in merging the documents with their citation counts. DOIs for each article can be obtained from the DBLP dump, then   used to query crossref's rest API to obtain  an approximate of the citation count. Of the 35,391 articles, citation counts were retrieved for 34,015 of them.

\begin{figure}[!t]
\centering
\includegraphics[scale=0.5]{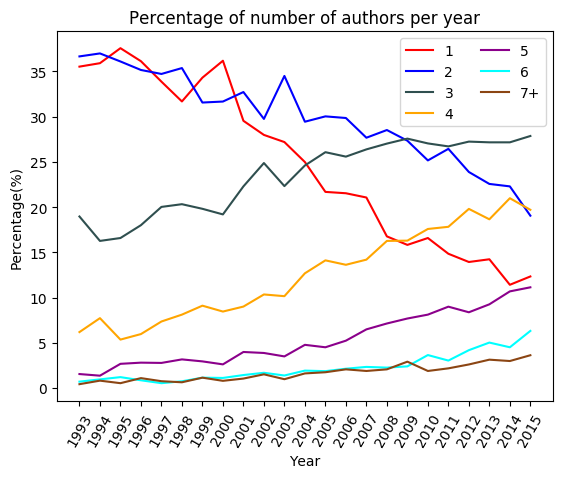}
\caption{Percentage of \# of authors co-authoring a paper in a year.}
\label{fig:coauthor_percent}
\end{figure}

\section{Results}
Our results with this data will recommend  $\mathit{PR}_W$, for three reasons. Firstly,
as shown in \tion{results}, an increasing trend in the SE literature is SE collaborations between multiple authors. Further,
those multiple-author papers are increasing prominent (cited most often). Hence, it is appropriate to switch from metrics like the h-index
to other metrics that reward collaboration like $\mathit{PR}_W$.

Secondly, as shown in \tion{results2},  $\mathit{PR}_W$ agrees with the most with all the other metrics studies here.  
Hence, if we used anything other than  $\mathit{PR}_W$,
then that would lead to more future disagreements about what ranking metric to use.

Thirdly,
as shown in \tion{results3},  $\mathit{PR}_W$ is numerical stable; i.e. perturbations in its derivation  have little impact
on its rankings.

\subsection{Results \#1: Why Focus on Collaboration?}
\label{sect:results}



\begin{table}[!t]
\centering
\scriptsize
\begin{tabular}{|l|P{0.65cm}|P{0.65cm}|P{0.65cm}|P{0.65cm}|P{0.65cm}|P{0.65cm}|P{0.65cm}|}\hline
\textbf{Year} & \textbf{1} & \textbf{2} & \textbf{3} & \textbf{4} & \textbf{5} & \textbf{6} & \textbf{7+}\\ \hline
1992 & \rA 0.08 & \mc{5}{\rB 0.24 \h 0.20 \h 0.32 \h 0.12 \h 0.20} & \rC 5.48  \\
1993 & \rA 0 & \mc{2}{\rB 0.13 \h 0.17} & \mc{2}{\rC 0.21 \h 0.29} & \rB 0.13 & -\\ 
1994 & \rA 0 & \mc{4}{\rB 0.17 \h 0.22 \h 0.35 \h 0.17} & \rC 0.65 & \rB 0.26 \\
1995 & \rA 0.05 & \mc{5}{\rB 0.23 \h 0.27 \h 0.32 \h 0.55 \h 0.18} & \rC 1.68 \\
1996 & \rA 0 & \mc{3}{\rB 0.10 \h 0.10 \h 0.14} & \rC 0.24 & \rB 0.05 & \rD 0.57 \\
1997 & \mc{5}{\rA 0.05 \h 0.10 \h 0.10 \h 0.15 \h 0.25} & \mc{2}{\rB 1.20 \h 0.50}\\
1998 & \rA 0.05 & \mc{6}{\rB 0.26 \h 0.26 \h 0.16 \h 0.21 \h 0.42 \h 0.21} \\
1999 & \rA 0 & \mc{5}{\rB 0.11 \h 0.11 \h 0.11 \h 0.06 \h 0.22} & -\\
2000 & \rA 0 & \mc{6}{\rB 0.18 \h 0.18 \h 0.18 \h 0.06 \h 0.12 \h 0.18}\\
2001 & \rA 0 & \rB 0.19 & \mc{5}{\rC 0.38 \h 0.38 \h 0.25 \h 0.25 \h 0.38} \\
2002 & \rA 0.07 & \mc{4}{\rB 0.33 \h 0.33 \h 0.47 \h 0.27} & \mc{2}{\rC 0.67 \h 0.53}\\
2003 & \rA 0.07 & \mc{6}{\rB 0.43 \h 0.43 \h 0.71 \h 0.57 \h 0.35 \h 0.58}\\
2004 & \rA 0.08 & \rB 0.23 & \mc{5}{\rC 0.38 \h 0.38 \h 0.54 \h 0.46 \h 0.71}\\
2005 & \rA 0.08 & \mc{6}{\rB 0.33 \h 0.33 \h 0.33 \h 0.33 \h 0.17 \h 0.33}\\
2006 & \rA 0 & \mc{2}{\rB 0.36 \h 0.36} & \mc{2}{\rC 0.45 \h 0.55} & \mc{2}{\rB 0.18 \h 0.27}\\
2007 & \rA 0.10 & \mc{2}{\rB 0.50 \h 0.60} & \mc{2}{\rC 0.70 \h 0.70} & \mc{2}{\rB 0.60 \h 0.50}\\
2008 & \rA 0.11 & \mc{4}{\rB 0.56 \h 0.78 \h 0.56 \h 0.67} & \mc{2}{\rC 0.89 \h 1.21}\\
2009 & \rA 0 & \rB 0.38 & \mc{4}{\rC 0.63 \h 0.75 \h 0.63 \h 0.75} & \rB 0.38\\
2010 & \rA 0.14 & \rB 0.57 & \rC 0.71 & \mc{4}{\rD 0.86 \h 1.00 \h 0.71 \h 1.00}\\
2011 & \rA 0 & \rB 0.67 & \mc{5}{\rC 0.83 \h 1.00 \h 1.00 \h 0.83 \h 1.00}\\
2012 & \rA 0 & \rB 0.60 & \rC 0.80 & \mc{4}{\rD 1.00 \h 1.40 \h 1.00 \h 1.20}\\
2013 & \rA 0 & \mc{2}{\rB 0.75 \h 0.75} & \mc{4}{\rC 1.00 \h 1.00 \h 1.00 \h 1.00}\\
2014 & \rA 0 & \rB 0.67 & \mc{4}{\rC 1.00 \h 1.00 \h 1.33 \h 1.00} & \rD 1.67\\
2015 & \rA 0 & \mc{2}{\rB 0.50 \h 0.50} & \mc{4}{\rC 1.00 \h 1.00 \h 1.00 \h 1.00}\\\hline
\end{tabular}

\begin{tabular}{cccc}
\\
\legend{rc1}{Rank 1} & \legend{rc2}{Rank 2} & \legend{rc3}{Rank 3} & \legend{rc4}{Rank 4}
\end{tabular}

\caption{Median value of average cites per year for articles with different number of coauthors. Cells with the same color have similar values
(as judged by an A12 test). In any row, cells with one color
have a different median value to cells of any other color in that row.  
Pink, green blue and white denote groups of cells with values that are lowest, not-so-low, higher an highest (respectively)    for any one year.}
\label{tab:authors_vs_cites}
\end{table}

\fref{coauthor_percent} shows the number of authors per paper within our corpus fro 1993 to 2016.
In that figure, we observe that:
\bi
\item
The number of
single author papers in SE has plummeted from 35\% (in 1993) to not
much more than 10\% (in 2015). 
\item
Similarly, over the same period, the number of double author papers has nearly halved.
\item
On the other hand, as observed by the positive trends in the results of all other curves, the number of paper with 3 or more authors has been steadily increasing.
\ei
So, clearly, the conclusion from \fref{coauthor_percent} is that the SE community is collaborating more. But does that collaboration lead to more prominent  papers? To answer that question, we turn to \tref{authors_vs_cites}:
\bi
\item The rows of that table show years from 1992 to 2015.
\item The columns of that table divide the results into those
from 1,2,3,4,5,6,7+ authors. 
\item Cells show  median  citations per paper in that year.
\item Cells with similar values have the same color, where ``similar'' is
is determined by a  non-parametric effect size test\footnote{Given two lists $x,y$ of length $m,n$
for each item in $u\in x$, count how many values $v\in y$    are  greater $g=\#(u>v)$ or equal $e=\#(u==v)$. According to  Vargha and Delaney~\cite{vd00}, if  $(g - e/2)/(m*n)<0.56$ then the two lists differ by a trivially small effect. } 
\item In each row, cells with similar values are grouped together and
their median is compared to other groups in that row. Cells
colored pink have lowest citations while cells colored green,blue and
white have successively higher citation counts.
\ei
Two important observations from    \tref{authors_vs_cites} are:
\bi
\item
All the single author papers  
are colored pink; i.e. they have the lowest citations of any group of papers.
\item
As we read down the rows, we can see an increasing frequency of blue and white cells for papers with three or more citations.  
\ei
Based on these two observations, we note that  if authors want their work read and cited, then single author papers should be deprecated.
Accordingly, we assert that  metrics that rank SE scholars should
include some measure of collaboration between authors (e.g.
as done in $\mathit{PR}_W$).






\begin{table}[!t]
\centering
\footnotesize
\begin{tabular}{l|c|c|c|c|c|c}
\cline{2-7}
 & \textbf{Infl} & \textbf{CoA} & \textbf{Harm} & \textbf{Frac} & \textbf{PR} & \multicolumn{1}{c|}{\textbf{$\text{PR}_\text{publ}$}} \\ \hline
\multicolumn{1}{|l|}{\textbf{Infl}} &  & 67 & 74 & 74 & 80 & \multicolumn{1}{c|}{79} \\ \hline
\multicolumn{1}{|l|}{\textbf{CoA}} & 67 &  & 59 & 71 & 72 & \multicolumn{1}{c|}{74} \\ \hline
\multicolumn{1}{|l|}{\textbf{Harm}} & 74 & 59 &  & \cellcolor[HTML]{34FF34}78 & 79 & \multicolumn{1}{c|}{79} \\ \hline
\multicolumn{1}{|l|}{\textbf{Frac}} & 74 & 71 & 78 &  & 77 & \multicolumn{1}{c|}{78}\\ \hline
\multicolumn{1}{|l|}{\textbf{PR}} & \cellcolor[HTML]{34FF34}80 & 72 & \cellcolor[HTML]{34FF34}79 & 77 &  & \multicolumn{1}{c|}{98} \\ \hline
\multicolumn{1}{|l|}{\textbf{$\text{PR}_\text{publ}$}} & \cellcolor[HTML]{34FF34}80 & \cellcolor[HTML]{34FF34}74 & \cellcolor[HTML]{34FF34}79 & \cellcolor[HTML]{34FF34}78 & \multicolumn{1}{c|}{\cellcolor[HTML]{34FF34}98} &  \\ \hline
\multicolumn{1}{|l|}{\textbf{$\text{PR}_\text{cite}$}} & 79 & 73 & \cellcolor[HTML]{34FF34}79 & \cellcolor[HTML]{34FF34}78 &  \cellcolor[HTML]{34FF34}98 & \multicolumn{1}{c|}{\cellcolor[HTML]{34FF34}99} \\ \hline
\end{tabular}
\caption{Percentage of common authors in the top 1\% of different ranking schemes. The best overlap for each column  is shown in green.}
\label{tab:rankingOverlap}
\end{table}


\subsection{Results \#2: Ranking Agreements}\label{sect:results2}

Another reason to prefer $\mathit{PR}_W$ it has most agreement with all the other metrics.
\tref{rankingOverlap} studies the top 1\% most cited authors in our corpus (ie. 3540 authors).
The cells of \tref{rankingOverlap} shows how often an author was ranked into group $X$ using
two different metrics. The green cells of that table show results of maximum overlap between the rankings generated by a column's metric to a row's metric.

The key observations from \tref{rankingOverlap} are:
\bi
\item The metrics which most agree with the other metrics are two $\mathit{PR}_W$ variants $\mathit{PR}_\mathit{cite}$ and 
$\mathit{PR}_\mathit{pub}$. 
\item When green cells appear in other rows, they always show a match that equals one of the  $\mathit{PR}_W$ rows.
\item
Hence, when selecting a metric that offers most of the same rankings as anything else, we recommend either of the $\mathit{PR}_W$ metrics
since this will lead to least future debates about the merits of alternate metrics.
\ei
Further to the last point, for pragmatic reasons,   we recommend  $\mathit{PR}_\mathit{cite}$  since that is closest to current practice (that is based on citation counts). Hence, $\mathit{PR}_\mathit{cite}$   may be least
disruptive to current career paths (therefore more palatable to more academics).

\subsection{Results \#3: Numerical Stability}\label{sect:results3}

Another observation we make $\mathit{PR}_W$ is that  there is a 98\% overlap between page rank (\textbf{PR}) and the two weighted page rank measures ($\mathit{PR}_{publ}$ and $\mathit{PR}_{cite}$). This suggests that the ranks generated with this measure are insensitive to the derivation
parameters of that metric.  Note that such insensitivity is a highly desirable property for a ranking metric since it means that the reported ranks are not effected by minor decisions within the calculation of that metric. We say that such a metric is {\em numerically stable}.

\begin{figure}[!t]
\includegraphics[scale=0.45]{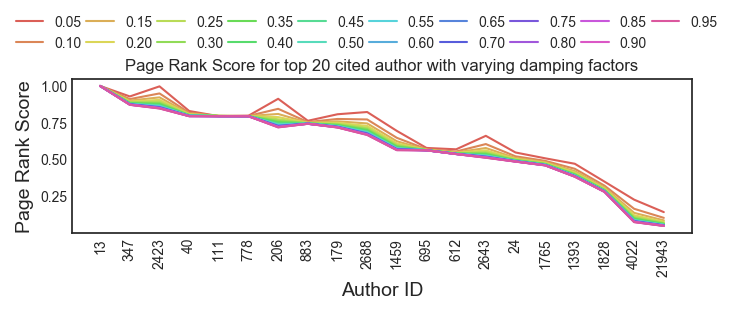}
\caption{$\mathit{PR}_W$ scores of top 20 most-cited authors for  $0 \le \theta \le 1$.}\label{fig:damp_pr} 
\end{figure}

To test is $\mathit{PR}_W$ is numerically stable we performed a perturbation study on Equation ~\ref{eq:pr} and ~\ref{eq:pr_w}.
Figure~\ref{fig:damp_pr} visualizes
the score of our 
top 20 most-cited authors using $\mathit{PR}$ and $\mathit{PR}_W$ for  $0 \le \theta \le 1$. This is figure shows results from
$\mathit{PR}$, $\mathit{PR}_{publ}$ and $\mathit{PR}_{cite}$

The key observation from Figure~\ref{fig:damp_pr} is that
the   relative ordering of the ranks are very similar. We can observe small changes at lower values($< 0.15$) of $\theta$ but at later values the ranks are almost similar. 

\section{Related Work}
\label{sec:rel_work}

The most similar recent study to this work was performed by Fernandes in 2014 who studies authorship trends in SE~\cite{fernandes2014authorship}. That study collected 70.000 papers entries from DBLP for 122 conferences and journals, for the between 1971–2012 and process several bibliometric indicators like CoA, Infl, Frac, Harm. He empirically shows trends like a) the number of authors of articles in software engineering is increasing on average around 0.40 authors/decade; b) Until 1980, the majority of the articles have had single authors, while more than half of the recent articles(2000 - 2012) have 3 to 4 co-authors associated with them. 
This study had two limitations a) It fails to address the impact of collaboration on SE which is a major part of SE which can be seen from \fref{coauthor_percent}. b) The study is not repeatable as data and algorithms are not made publically available.

In their 2014 Elsevier newsletter, Plume and van Weijen studied contribution of authors to SE research~\cite{plumepublish}. They show that there has been no increase in contribution per active author over the last decade. They theorize that authors use their authorship potential to become more collaborative in the way they work. 
We can empirically see from \fref{coauthor_percent} that there has been a rise in collaboration over the years in SE. This can be attributed to the reward of more citations from collaboration(as seen in \tref{authors_vs_cites}).


\begin{table*}[!t]
\begin{center}

\resizebox{.6\textwidth}{!}{%
\begin{tabular}{|l|l|l|l|l|l|}\hline 
 \textbf{Object-Oriented}  & \textbf{Testing} & \textbf{Source Code} & \textbf{Architecture} & \textbf{Modeling} & \textbf{Developer} \\\hline
    
   J Sheffield &  M Harman & D Poshyvanyk & S Apel & DL Moody & T Dyba \\
   PR Houser & A Acuri & R Oliveto & P Avgeriou & M Dumas & T Zimmermann \\
    B Doty & G Fraser & A de Lucia & C Kastner & M Chechik & T Dingsoyr \\
   Y Tian  & AM Memon & M di Penta & U Kulesza & S Uchitel & P Runeson \\
  L Lighty &  LC Briand & Y Gueheneuc & J White & C Ouyang & M Host \\
    \hline
    \end{tabular}}
    
    \vspace{5mm}
    
    \resizebox{.6\textwidth}{!}{%
 \begin{tabular}{|l|l|l|l|l|}\hline 
     \textbf{Program Analysis  } & \textbf{Requirements} & \textbf{Metrics} & \textbf{Applications} & \textbf{Performance}  \\
  \hline
    
    W Weimer & BA Kitchenham & T Menzies & R Buyya & C Chang   \\
    A Orso & T Gorschek & N Nagappan &  CAF De Rose & C Tsai   \\
    MD Ernst & P Brereton & T Zimmermann & RN Calheiros & W Hong   \\
    A Zeller & MP Velthius & MJ Shepperd & R Ranjan & T Chen  \\
    S Forrest & D Budgen & AE Hassan & A Beloglazov & C Yang  \\ \hline
    \end{tabular}}

 \end{center}
    \caption{The top 5 authors  seen in 11 SE topics since 2008. Scores  calculated
    using $\mathit{PR}_\mathit{cite}$. Topics determined as per \cite{mathewTrends17}. }
    \label{tab:topAuthorsTopicsRecent}
\end{table*}
\begin{table*}[!t]
\begin{center}

\resizebox{.6\textwidth}{!}{%
\begin{tabular}{|l|l|l|l|l|l|}\hline 
 \textbf{Object-Oriented}  & \textbf{Testing} & \textbf{Source Code} & \textbf{Architecture} & \textbf{Modeling} & \textbf{Developer} \\\hline
    
   VR Basili & G Rothermel & F Antioniol & N Medvidovic & GJ Holzmann & JD Herbsleb \\
   MV Zelkowitz & MJ Harrold &  A DeLucia & RN Taylor & J Kramer & BA Kitchenham \\
   O Laitenberger & M Harman & A Marcus & D Garlan & LC Briand & A Mockus \\
   RL Glass & J Offutt & D Poshyvank & J Bosch & S Uchitel & T Dyba \\
   F Shull & LC Briand & GC Murphy & J OckerBloom & B Nuseibeh & T Zimmermann \\
    \hline
    \end{tabular}}
    
    \vspace{5mm}
    
    \resizebox{.6\textwidth}{!}{%
 \begin{tabular}{|l|l|l|l|l|}\hline 
     \textbf{Program Analysis  } & \textbf{Requirements} & \textbf{Metrics} & \textbf{Applications} & \textbf{Performance}  \\
  \hline
    
    MD Ernst & CF Kemerer & LC Briand & R Buyya & TMJ Fruchterman\\
    A Zeller & SR Chidamber & MJ Shepperd & B Benatallah & EM Reingold \\
    A Orso & BA Kitchenham & NE Fenton & M Dumas & C Chang \\
    MJ Harrold & BW Boehm & T Menzies & J Kalagnanam & T Chen\\
    K Sen & AI Anton & BA Kitchenham & H Chang & M Hwang\\ \hline
    \end{tabular}}

 \end{center}
    \caption{The top 5 authors  seen in 11 SE topics in the corpus. Scores  calculated
    using $\mathit{PR}_\mathit{cite}$. Topics determined as per \cite{mathewTrends17}. }
    \label{tab:topAuthorsTopicsAll}
\end{table*}

\section{Conclusion}
\label{sec:conc}

The above discussion recommends $\mathit{PR}_\mathit{cite}$ 
since it incorporates aspects of author collaboration
as well as more traditional citation counts. Also, this metric
has most agreement with other metrics which means that rankings
generated by this metric are less likely to be refuted by other metrics. Further, $\mathit{PR}_\mathit{cite}$ 
is numerically stable.

\tref{topAuthorsTopicsRecent} uses   $\mathit{PR}_\mathit{cite}$  to show the top authors from various SE topics, in the last decade (since 2009). The topics of that figure were discovered by a
text mining methods called Latent Dirichlet Allocation that automatically discover groups of terms that cover most documents.
For more details on that analysis, see~\cite{mathew2017TSE}. 

Note that tables like \tref{topAuthorsTopicsRecent} are not the goal of this research, Rather, our main point
is that if status is so important to SE scholars, then the methods used to assess that status need to be debated
by this community. Any proposal that SE scholars should be assessed using metric \textit{Weighted PageRank}($\mathit{PR}_W$) needs
to be carefully audited. More specifically, as done here, the analysis that recommends metrics $\mathit{PR}_W$ needs
to be automatic and repeatable. To this end, we offer all our data and scripts at \href{https://goo.gl/xnB6f3}{goo.gl/xnB6f3}. We strongly
encourage other researchers to  be as forthcoming with their assessment material.




%
 
\bibliographystyle{IEEEtran}
\bibliography{IEEEabrv,refs}



%








\end{document}